\documentclass[12pt]{iopart}

\usepackage{epsfig}
\usepackage{setstack}
\usepackage{iopams}

\begin{document}

\title{A Hamiltonian approach to quantum gravity}

\author{E V Stefanovich}

\address{2255 Showers Drive, Apt. 153, Mountain View, California 94040, USA}
\ead{eugene\_stefanovich@usa.net}

\begin{abstract}
We explore the idea that gravitational interaction can be described
by instantaneous inter-particle potentials. This idea is in full
accord with relativistic quantum theory. In particular, it resembles
the ``dressed particle'' approach to quantum electrodynamics.
Although the complete non-perturbative of this theory is yet
unknown, one can reasonably guess its form in low perturbation
orders and in the $(1/c)^2$ approximation. We suggest a relativistic
energy operator, which in the classical limit reduces to the
Einstein-Infeld-Hoffmann Hamiltonian for massive particles and
correctly describes the effects of gravity on photons, including the
light bending, the Shapiro delay, the gravitational time dilation
and the red shift. The causality of this approach is briefly
discussed.
\end{abstract}


\section{ Introduction} \label{sc:intro}

Formulation of the quantum theory of gravity is still a major
unresolved problem in theoretical physics. It is clear that direct
combination of quantum mechanics with general relativity is
impossible, because these disciplines treat space and time in very
different ways \cite{Isham, Macias}. Quantum theory is commonly
expressed in the Hamiltonian formalism, where positions of particles
are dynamical variables (Hermitian operators), but time is a
numerical parameter labeling reference frames. On the other hand, in
general relativity both time and position are coordinates in the
curved space-time manifold.  Various attempts to adjust quantum
mechanics to the requirements of general relativity have not
produced a viable theory yet.

 In this paper we will explore a different
approach in which gravity is described within the same Hamiltonian
formalism as ordinary relativistic quantum mechanics, without
introduction of the space-time curvature. We will assume that
quantum theory of gravity can be built in analogy with the most
successful theory of particle interactions - quantum electrodynamics
(QED). The ``dressed particle'' approach \cite{mybook} suggests that
field-based QED can be equivalently formulated as a theory of
particles (electrons, positrons, photons, etc) interacting via
instantaneous potentials that depend on particles' positions and
velocities and, in the general case, can change the number of
particles as well. We will assume that the same properties hold for
the (yet unknown) full non-perturbative ``dressed particle''
formulation of the gravitational theory. Due to the weakness of
gravitational interactions, we can also assume that the perturbative
expansion converges rapidly and that for most practical purposes the
second perturbation order in the coupling constant should be
sufficient. From the analogy with QED we can conclude that in this
order the processes of particle creation and annihilation are not
allowed and that interactions are well reproduced by two-particle
potentials (similar to the Darwin-Breit potential between charged
particles in QED). Three conditions will be used to guess the
functional form of these potentials. First, in the classical limit
and in the $(1/c)^2$ approximation this approach must reduce to the
well-known post-Newtonian approximation of general relativity, which
is widely used for calculations of relativistic gravitational
effects. Second, the theory must be relativistically invariant,
therefore the Poincar\'e commutation relations involving the
interacting Hamiltonian and boost operators should remain valid, at
least in the $(1/c)^2$ approximation. Third, the same Hamiltonian
should describe interactions between massive particles as well as
interactions of massive particles with photons. The Hamiltonian
fulfilling all these requirements is written in (\ref{eq:ham}), and
in the rest of the paper we are demonstrating that physical
predictions following from this Hamiltonian agree with available
experimental and observational data, such as dynamics of the Solar
system, light bending and propagation delay, gravitational red shift
and time dilation.

\section{Relativistic Hamiltonian dynamics} \label{sc:intro2}

The principles of relativity and quantum mechanics are most
naturally combined within Wigner-Dirac theory of unitary
representations of the Poincar\'e group \cite{Wigner_unit, Dirac,
book}. According to this theory, a full description of dynamics of
any isolated physical system requires construction of a
representation of the Poincar\'e Lie algebra by Hermitian operators
in the Hilbert space of states. Representatives of ten generators of
the algebra are identified with total observables of the system:
$\bi{P}$ (total linear momentum), $\bi{J}$ (total angular momentum),
$H$ (total energy), and $\bi{K}$ (for systems with zero spin this
observable corresponds to the product of the center-of-mass position
$\bi{R}$ and total energy: $\bi{K} = -1/2(H\bi{R} + \bi{R}H)$, so it
can be called the \emph{center of energy}). The commutators of
generators\footnote{Indices $a,b,c$ label Cartesian coordinates
($x,y,z$) of 3-vectors.}

\begin{eqnarray}
[J_a, P_b] &=& i\hbar \sum_{c=1}^3 \epsilon_{abc} P_c
\label{eq:5.50}
\\
\mbox{ } [J_a, J_b] &=& i\hbar \sum_{c=1}^3 \epsilon_{abc} J_c
\label{eq:5.51}
\\
\mbox{ } [J_a, K_b] &= &i\hbar \sum_{c=1}^3 \epsilon_{abc} K_c
\label{eq:5.52}
\\
\mbox{ } [P_a,P_b] &=&  [J_a,H] = [P_a, H] = 0 \label{eq:5.53}
\\
 \mbox{ } [K_a, K_b] &=& -i\frac{\hbar}{c^2} \sum_{c=1}^3 \epsilon_{abc}
 J_c
\label{eq:kikj}
\\
\mbox{ } [K_a, P_b] &=& -i\frac{\hbar}{c^2} H \delta_{ab} \label{eq:kipj}\\
 \mbox{ } [K_a, H] &=& -i\hbar P_a
\label{eq:kih}
\end{eqnarray}

\noindent  play two important roles. First, they determine whether
observables can be measured simultaneously. Second, they tell us how
 observables change with respect to inertial transformations of
observers. For example, the change of any observable $F$ with
respect to time translations is

\begin{eqnarray*}
F(t)  &=& e^{\frac{i}{\hbar} Ht} F e^{-\frac{i}{\hbar} Ht}  = F +
\frac{it}{\hbar} [H,F] - \frac{t^2}{2! \hbar^2} [H, [H,F]] + \ldots
\label{eq:5.61}
\end{eqnarray*}

 In the instant form of Dirac's
dynamics \cite{Dirac}, the interaction is encoded in the form of
generators $H$ and $\bi{K}$, which are different from their
non-interacting counterparts  $H_0$ and $\bi{K}_0$.

\begin{eqnarray}
 \bi{P} &=& \bi{P}_0  \label{eq:P}\\
\bi{J} &=& \bi{J}_0  \label{eq:J} \\
\bi{K} &=& \bi{K}_0 + \bi{Z} \label{eq:k-z}\\
H &=& H_0 + V. \label{eq:H}
\end{eqnarray}

\noindent The above relationships constitute the foundation of
relativistic quantum theories of electromagnetic and nuclear forces
\cite{book, mybook}. The "dressed particle" formalism
\cite{Shirokov4, Stefanovich_qft, mybook} allows one to eliminate
field variables from these theories and to formulate them in terms
of directly measurable quantities -- particle observables.

\section {Two-body systems} \label{ss:gravit-pot}

The general theory presented above has simple realization in the
case of two spinless particles with gravitational interaction. Let
us denote one-particle observables by small letters: momentum
$\bi{p}$, position $\bi{r}$, mass $m$, energy $ h = \sqrt{m^2c^4 +
p^2c^2} $, and center of energy $ \bi{k} = -h\bi{r}c^{-2} $.
 We will also switch to the
classical limit $\hbar \to 0$, where commutators of operators can be
replaced by Poisson brackets

\begin{eqnarray*}
 \lim_{\hbar \to 0}(-\frac{i}{\hbar})[F, G] = \{F,G\} \nonumber \\
\equiv \left(\frac{\partial F}{
\partial\bi{r}_1 } \cdot \frac{\partial G}{
\partial\bi{p}_1 }\right) - \left(\frac{\partial F}{ \partial\bi{p}_1 } \cdot
\frac{\partial G}{ \partial\bi{r}_1 }\right) + \left(\frac{\partial
F}{
\partial\bi{r}_2 } \cdot \frac{\partial G}{
\partial\bi{p}_2 }\right) - \left(\frac{\partial F}{ \partial\bi{p}_2 } \cdot
\frac{\partial G}{ \partial\bi{r}_2 }\right).
\end{eqnarray*}

\noindent In this limit, operators of observables commute, so their
order in products becomes irrelevant, and there is no unique method
to restore the original quantum theory from its classical limit.
However, this ambiguity affects only terms proportional to $\hbar$,
which are too small to be observable in most experiments.

For non-interacting particles the generators are simply sums of
one-particle terms

\begin{eqnarray*}
\bi{P}_0 &=& \bi{p}_1 + \bi{p}_2 \label{eq:p00} \\
\bi{J}_0 &=& [\bi{r}_1 \times \bi{p}_1] +
 [\bi{r}_2 \times \bi{p}_2] \label{eq:j00} \\
H_0 &=& h_1 + h_2 \label{eq:ham0}\\
\bi{K}_0 &=& \bi{k}_1 + \bi{k}_2.  \label{eq:boost0}
\end{eqnarray*}

\noindent  In order to fulfil three requirements from Introduction,
let us consider the following interacting generators

\begin{eqnarray}
H &=& H_0  - \frac{Gh_1h_2}{c^4r} - \frac{Gh_2p_1^2}{h_1c^2r} -
\frac{Gh_1p_2^2}{h_2c^2r} + \frac{7G(\bi{p}_1 \cdot
\bi{p}_2)}{2c^2r} + \frac{G(\bi{p}_1 \cdot
\bi{r})(\bi{p}_2 \cdot \bi{r})}{2c^2r^3} \nonumber \\
&+& \frac{G^2m_1m_2(m_1+m_2)}{2c^2r^2} \label{eq:ham}\\
\bi{K} &=& \bi{K}_0 + \frac{Gh_1h_2 (\bi{r}_1 + \bi{r}_2)}{2c^6 r}
\label{eq:boost}
\end{eqnarray}

\noindent where $\bi{r} \equiv \bi{r}_1 - \bi{r}_2$. These
expressions will be considered as second order (1PN) approximations
with respect to the smallness parameter $(1/c)$.\footnote{note that
$h_i$ are of order $c^2$ and $\bi{K}_0$ is of order $(1/c)^2$} Then
using canonical Poisson brackets between positions and momenta, one
can show by a straightforward calculation that (\ref{eq:5.50}) -
(\ref{eq:kih}) are satisfied in this order.

Consider two massive bodies whose velocities are small in comparison
with the speed of light $(p \ll mc)$. Then replacing their energies
in (\ref{eq:ham}) by the expansion

\begin{eqnarray*}
 h &\approx& mc^2 + \frac{p^2}{2m} - \frac{p^4}{8m^3c^2} +
\ldots
\end{eqnarray*}

\noindent we see that $H$ coincides with the
Einstein-Infeld-Hoffmann Hamiltonian \cite{EIH}

\begin{eqnarray}
H &=& m_1c^2 + m_2c^2 +\frac{p_1^2}{2m_1} + \frac{p_2^2}{2m_2} -
\frac{Gm_1m_2}{r} - \frac{p_1^4}{8m_1^3c^2} -
\frac{p_2^4}{8m_2^3c^2} \nonumber \\
&-& \frac{3Gm_2p_1^2}{2m_1c^2r} - \frac{3Gm_1p_2^2}{2m_2c^2r} +
\frac{7G(\bi{p}_1 \cdot \bi{p}_2)}{2c^2r} + \frac{G(\bi{p}_1 \cdot
\bi{r})(\bi{p}_2 \cdot \bi{r})}{2c^2r^3} \nonumber \\
&+& \frac{G^2m_1m_2(m_1+m_2)}{2c^2r^2}. \label{eq:final-h}
\end{eqnarray}

\noindent which is usually obtained in the 1PN approximation to
general relativity (see, e.g., \S 106 in \cite{Landau2}). This
Hamiltonian correctly describes precession of the Mercury's orbit
\cite{Landau2, Damour} and dynamics of the Sun-Earth-Moon system
\cite{Damour-96, Nordtvedt, Turyshev-03}.

\section { Photons} \label{ss:photon-ham}

It is interesting to note that Hamiltonian (\ref{eq:ham}) can
describe the interaction of massive bodies with light as well. We
will consider the case when the massive body 2 is very heavy (e.g.,
Sun), so that photon's momentum satisfies inequality $p_1 \ll m_2c$.
Then in the center-of-mass frame ($\bi{p}_1 = -\bi{p}_2 \equiv
\bi{p}$) we can take the limit $m_1 \to 0$, replace $h_1 \to pc$, $
h_2 \to m_2c^2$,
 and ignore the inconsequential rest energy $m_2c^2$ of the massive
 body. Then from (\ref{eq:ham})
we obtain a Hamiltonian accurate to the order $(1/c)$

\begin{eqnarray}
H &=&  pc  - \frac{2Gm_2p}{cr} \label{eq:phot-pot}
\end{eqnarray}

\noindent which can be used to evaluate the motion of photons in the
Sun's gravitational field. The time derivative of the photon's
momentum can be found from the first Hamilton's equation

\begin{eqnarray}
\frac{\rmd \bi{p}}{\rmd t}  = -\frac{\partial H}{\partial \bi{r}} =
-\frac{2G m_2 p \bi{r}}{cr^3}. \label{eq:phot-dqdt}
\end{eqnarray}

\begin{figure}
\epsfig {file=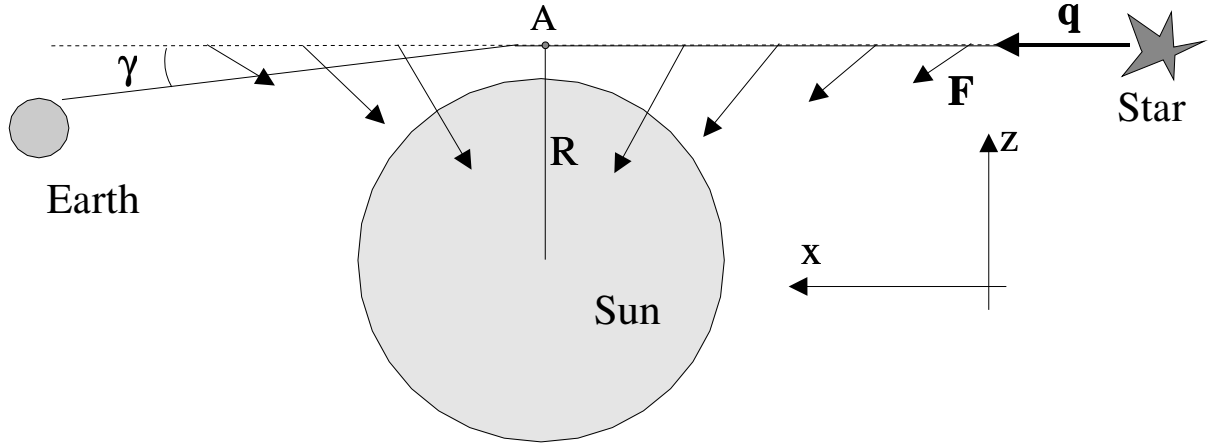} \caption{ Light bending by  Sun's gravity.}
\label{fig:1}
\end{figure}

\noindent In the zeroth approximation we can assume that the photon
moves with the speed $c$ along the straight line ($x = ct$) and the
\emph{impact parameter} is $ R$ (see figure \ref{fig:1}). Then  the
accumulated momentum in the $z$-direction is obtained by integrating
the $z$-component of (\ref{eq:phot-dqdt})

\begin{eqnarray*}
\Delta p_{z}
  \approx -\int_{-\infty}^{\infty} \frac{2G m_2 pR \rmd t}{c(R^2 + c^2t^2)^{3/2}}
  = -\frac{4G m_2 p}{c^2R}.
\end{eqnarray*}

\noindent The deflection angle

\begin{eqnarray*}
\gamma &\approx& \tan \gamma  = \frac{|\Delta p_{z}|}{p} =
\frac{4Gm_2}{c^2R}
\end{eqnarray*}

\noindent  coincides with the observed bending of starlight by the
Sun's gravity \cite{Will-review}.

The second Hamilton's equation

\begin{eqnarray*}
\frac{\rmd \bi{r}}{\rmd t} &=& \frac{\partial H}{\partial \bi{p}} =
\frac{\bi{p}}{p}\Bigl(c - \frac{2G m_2 }{cr}\Bigr)
\label{eq:phot-drdt}
\end{eqnarray*}

\noindent  can be interpreted
 as gravitational reduction of the speed of light. This means that in the presence of gravity it takes
photons an extra time to travel the same path. Let us find the time
delay for a photon traveling from the point $A$ on the Sun's surface
(see fig. \ref{fig:1}) to the observer on Earth. Denoting $d$ the
distance Sun - Earth and taking into account that $R \ll d$ we
obtain

\begin{eqnarray*}
\Delta t &\approx& \frac{1}{c} \int _{0}^{d/c} \frac{ 2Gm_2 \rmd t}{
c(R^2 + c^2 t^2)^{1/2}} = \frac{2Gm_2}{c^3} \log(\frac{2d}{R})
\end{eqnarray*}

\noindent  which agrees with the leading general-relativistic
contribution to the propagation delay of radar signals near the Sun
\cite{Will-review}.

\section {Red shift and time dilation} \label{sc:dilation}

Electromagnetic radiation is normally emitted in  transitions
between two energy levels ($E_i >E_f$) in a multiparticle system,
such as atom, molecule, nucleus, etc. A source which is far from
massive bodies emits photons with frequency

\begin{eqnarray}
\nu(\infty)  &=& \frac{2 \pi}{\hbar} (E_i - E_f). \label{eq:freq}
\end{eqnarray}

\noindent In states $i$ and $f$ the source has different masses $m_i
= E_i/c^2$ and $m_f = E_f/c^2$, respectively. This means that
gravitational attraction is different in these states. The initial
and final total energies of the system "stationary source + Earth"
can be obtained from (\ref{eq:final-h})

\begin{eqnarray*}
\mathcal{E}_i &\approx& Mc^2 + E_i - \frac{GME_i}{Rc^2} \\
\mathcal{E}_f &\approx& Mc^2 + E_f - \frac{GME_f}{Rc^2}
\end{eqnarray*}

\noindent where $M$ is the Earth's mass and $R$ is the Earth's
radius. Then the frequency of radiation emitted by the source on
Earth is reduced in comparison with (\ref{eq:freq})

\begin{eqnarray*}
\nu(R)  &=& \frac{2 \pi}{\hbar} (\mathcal{E}_i - \mathcal{E}_f)
  \approx  \nu(\infty)(1 - \frac{GM}{Rc^2}). \label{eq:nu-phi}
\end{eqnarray*}

\noindent Gravitational red shift experiments \cite{Will-review}
confirmed this formula to a high precision. They are usually
performed by using identical systems (e.g, $^{57}$Fe nuclei in
M\"ossbauer experiments) as both the source and the detector of
radiation. If the source and the detector are at different
elevations (different gravitational potentials), then the mismatch
in their energy level separations makes the resonant absorption
impossible.

Note that during its travel from the source to the detector, the
photon's kinetic energy ($cp$) varies according to
(\ref{eq:phot-pot}). However, when the photon gets absorbed by the
detector it disappears completely, so its \emph{total} energy
(kinetic plus potential) gets transferred to the detector rather
than the kinetic energy alone. The photon's total energy (and
frequency) remains constant during its travel, so the attraction of
photons to massive bodies (\ref{eq:phot-pot}) does not play any role
in the gravitational red shift \cite{Okun}.

Gravitational time dilation experiments \cite{Will-review} are
fundamentally similar to red shift experiments discussed above,
because (atomic) clocks are also non-stationary quantum systems
whose oscillation frequency is proportional to the energy level
separation. The major difference is that low frequencies (e.g.,
those in the radio or microwave spectrum) are involved, so that
oscillations can be reliably counted and mapped to the time domain.
Even in most accurate time dilation and red shift measurements only
the effect of the low order Newtonian potential has been verified.
Post-Newtonian corrections in (\ref{eq:final-h}) can be measured in
future space missions \cite{Lammerzahl, Iorio, Turyshev}.

\section {Discussion}

We suggested a simple two-body approximation (\ref{eq:ham}) for the
 Hamiltonian of quantum gravity. For massive bodies in the 1PN approximation it
reduces to the well-known Einstein-Infeld-Hoffmann expression
\cite{EIH}, which successfully describes major observational data.
The same Hamiltonian (\ref{eq:ham}), when applied to the action of
gravity on massless particles (photons), correctly reproduces
standard general-relativistic results for the light deflection and
the radar echo time delay. This Hamiltonian description of gravity
can be extended to higher orders in $(1/c)$, so that Poincar\'e
commutation relations remain valid and detailed description of
dynamics of binary pulsar systems becomes possible, including
effects of radiation reaction \cite{PhysRevD.62.021501, Andrade,
Damour-07, Blanchet} and spin \cite{PhysRevD.63.044006, Porto}.

In general relativity, post-Newtonian Hamiltonians with
instantaneous interactions are considered as approximations to the
exact Einstein's equation for the curvature tensor. It is generally
believed that instantaneous potentials should lead to causality
paradoxes \cite{Russo} and are, therefore, unacceptable. However,
the paradoxes occur only if one uses Lorentz formulas to transform
space-time coordinates of events between different reference frames.
It was shown in \cite{Stefanovich_Mink, mybook} that due to the
presence of interaction terms in the boost generator
(\ref{eq:boost}), Lorentz transformations for events in interacting
systems of particles must be modified. This modification ensures
that the chronological order of events remains the same in all
frames of reference. The generally accepted idea of the retarded
propagation of gravity is not supported by any experimental data
\cite{Flandern, Flandern-Vigier}. Recent claims about measurements
of the finite speed of gravity \cite{Fomalont-Kopeikin, Kopeikin}
were challenged in a number of publications (see section 3.4.3 in
\cite{Will-review}). Therefore, at this point, there are no
indications that the Hamiltonian approach has any irreconcilable
contradictions with experiment or important theoretical principles.
It is plausible that the exact theory of gravity (whose $(1/c)^2$
approximation was examined in this work) may have the form
(\ref{eq:P}) - (\ref{eq:H}), where interactions $V$ and $\bi{Z}$ are
sums of instantaneous distance- and velocity-dependent potentials.

\ack

I would like to thank Juan R. Gonz\'alez-\'Alvarez and Eugene
Shubert for helpful online discussions. I am also thankful to two
anonymous referees whose critical comments lead to radical
improvements in this work.

\section*{References}


\end{document}